\documentclass[aps,prb,twocolumn,floats]{revtex4-2}
\usepackage{epsfig}
\usepackage{amsmath, amssymb}
\usepackage{graphicx}
\usepackage{soul}
\usepackage{color}

\begin{document}
\title{Edge state behavior in a Su-Schrieffer-Heeger like model with periodically modulated hopping}
\author{Satyaki Kar}
\affiliation{AKPC Mahavidyalaya, Bengai, West Bengal -712611, India}
\begin{abstract}
  Su-Schrieffer-Heeger (SSH) model is one of the simplest models to show topological end/edge states and the existence of Majorana fermions. Here we consider a SSH like model both in one and two dimensions where a nearest neighbor hopping features spatially periodic modulations. In the 1D chain,  we witness appearance of new in-gap end states apart from a pair of Majorana zero modes (MZM) when the hopping periodicity go beyond two lattice spacings. The pair of MZMs, that appear in the topological regime, characterize the end modes each existing in either end of the chain. These, however, crossover to both-end end modes for small hopping modulation strength in a finite chain. Contrarily in a 2D SSH model with symmetric hopping that we consider, both non-zero and zero energy topological states appear in a finite square lattice even with a simple staggered hopping, though the zero energy modes disappear in a ribbon configuration. Apart from edge modes, the 2D system also features corner modes as well as modes with satellite peaks distributed non-randomly within the lattice. In both the dimensions, an increase in the periodicity of hopping modulation causes the zero energy Majorana modes to become available for either sign of the modulation. But interestingly with different periodicity for hopping modulations in the two directions, the zero energy modes in a 2D model become rarer and does not appear for all strength and sign of the modulation.
\end{abstract}
\maketitle
\section{Introduction}
Quantum computers, that are much more efficient and faster than the supercomputers, enjoy topological protections when their quantum gates operate on non-Abelian quasiparticles called anyons\cite{nayak,elliot,beenakker}. Thus they constitute the key to the success of topological quantum computation. Majorana fermions\cite{elliot,beenakker,wilczek} (MF) are one noteworthy example of such quasiparticles proposed/found in exotic condensed matter systems. For being their own antiparticles, they are elusive as elementary particles in the high energy physics systems. But planar Josephson junctions\cite{ren}, spin-orbit coupled quantum wires\cite{mourik}, or topological superconductors\cite{leijnse} give us the opportunity to rather easily find them as emergent quasiparticle excitations. There a fermionic state that can be thought of as superposition of two MFs, is broken down to two spatially separated Majorana states and due to their non abelian exchange statistics, they lead to decoherence-free quantum information processing\cite{meyer}. Due to particle-hole symmetry MFs appear at zero energies. And such zero modes are mostly boundary states that shapes the topological transport in these systems.
There were proposals to create MFs in vortices of topological insulators\cite{ashwin}, carbon nanotubes\cite{loss}, chains of quantum dots\cite{jdsau} etc. Experimental findings also back up such claims\cite{mourik,tewari}.
Majorana zero modes (MZM) have also been proposed in strongly correlated systems like quantum Hall states or Kitaev-type spin models\cite{fu}. But in this work we are rather interested in free fermionic models which also feature such exotic excitations.

A simple Kitaev chain\cite{kitaev}, $i.e.,$ a nearest neighbor tight binding chain with spinless $p$-wave superconducting pairing has been proposed to host MFs at its two ends. Similarly, a Su-Schrieffer-Heeger (SSH) model, that can correspond to a polyacetylene chain, is a 1D tight binding model with staggered hopping modulation\cite{ssh} and it can demonstrate charge fractionalizations\cite{meier,ziani} and the existence of zero energy end states which are Majorana bound states (MBS) in the topological regime. {Though of different origins, one can find many similarities between these two Majorana hopping models.}

A chiral symmetric SSH chain shows topological phase for a negative value of the hopping modulation ({in fact, it depends on whether a strong or a weak hopping/bond is in excess in the finite lattice}) and  the SSH chain dispersion plot clearly shows the topological difference between positive and negative modulations. Thus it implies a topological phase transition whenever the hopping modulation changes its sign. Hence, interesting becomes the case where MBS gets redistributed/modified due to a periodic variation of the hopping modulation that goes beyond simple staggered hopping of the SSH chain. More specifically, in this paper we will analyze a SSH like chain which has a multi-sublattice structure (unlike the pure SSH chain which is a lattice with a 2 atom basis) depending on the periodicity of the hopping. Similar study has been done by Lin $et.~al.$\cite{lin} which discussed topological superconductivity in periodically modulated Majorana chains and analyzed spectral and topological response under phase variation of the modulated hopping. Our emphasis, on the other hand, is on considering single-phase mode of variation of the hopping and discuss the alteration of end modes including MBS as a function of hopping strength as well as the modulation periodicity.

Furthermore, we extend our study to two dimensions where we probe the boundary modes under periodic variation of hopping along both directions, say $\hat x$ and $\hat y$. With symmetric SSH type hopping along the two directions, topological phases are reported in these models in the absence of any Berry curvature\cite{liu}. We probe the features of the topological states as the hopping modulation is varied periodically. In the following section, we summarize our results, mention its novelty and discuss on possible future work directions.

\section{Formulation}
\subsection{1D Chain}
We first consider a finite SSH like chain with open boundary condition which can be expressed, both in Complex and Majorana fermion representation, as
  \begin{align}
    H&=\sum_{i=1}^{L-1}(t+\delta_{i}) c_i^\dagger c_{i+1} + h.c.\nonumber\\
    &=\sum_{i=1}^{L-1}i(t+\delta_{i}) [\gamma_{i,A}\gamma_{i+1,B}+\gamma_{i+1,A}\gamma_{i,B}]
    \label{ssh1d}
  \end{align}
  with $\gamma_{iA}(\gamma_{iB})=\frac{1(i)}{\sqrt{2}}[c_{i}^\dagger+(-)c_{i}]$ denoting the Majorana operators satisfying $\gamma^\dagger=\gamma$.
{ Notice that here the Majorana formulation is important as it shows the Hamiltonian to be decoupled into that of two Majorana hopping chains ($e.g.$, see Eq.\ref{decoupled} for $\theta=\pi$ later on) which are independent of each other\cite{fu}}.
  Here the periodic modulation in hopping strength is given by $\delta_i=\Delta cos[(i-1)\theta]$ and we study the system for different values of angle $\theta$. Notice that there is no hopping modulation for $\Delta=0$, irrespective of any $\theta$ or for $\theta=0$, irrespective of any $\Delta$. With nonzero $\Delta$ and $\theta$, nontrivial topological phases can appear in the system with the presence of MFs at the boundaries. Original SSH model can be retrieved with $\theta=\pi$ which gives topological end states for $\Delta/t\le0$ and trivial gapped phases otherwise. Notice that for an open chain with even number of sites, there is a number difference between strong and weak bonds and that gives the spectral difference between positive and negative $\Delta$ as seen in Fig.1. For $\theta=\pi/2$, we don't see any such spectral difference caused by the sign of $\Delta$, though it again appears for $\theta=\pi/4$. Thus this sign dependence rely on the value of $\theta$ chosen. For $\theta=\pi,~\pi/4$, there are different numbers of maximum ($t+\Delta$) and minimum ($t-\Delta$) hopping terms in the Hamiltonian causing non-identical spectral results for positive and negative $\Delta$ but for $\theta=\pi/2$, there are equal number of maximum and minimum hopping strengths and hence no spectral difference due to the sign of $\Delta$ occurs.

  Now we have to understand the topological nontriviality of the system. A SSH chain possess two zero energy Majorana modes $\gamma^{(1),(2)}$ satisfying $[H,\gamma^{(1),(2)}]=0$. In general for a Majorana Hamiltonian $H=\sum_{ij}A_{ij}i\gamma_i\gamma_j$, a zero mode $\sum_i\lambda_i\gamma_i$ requires\cite{fu,sarma}
  \begin{align}
    [\sum_i\lambda_i\gamma_i,\sum_{jk}iA_{jk}\gamma_j\gamma_k]=0~~~~{\rm or}~~~~\sum_iA_{ki}\lambda_i=0 .
    \label{0mode}
    \end{align}
 For example for $\theta=\pi$ in the Hamiltonian \ref{ssh1d}, there is a two sublattice structure in the model and the Hamiltonian constitutes two Majorana chains that do not talk to each other:
 \begin{align}
   H&=i[(t+\Delta)\gamma_{1,A}\gamma_{2,B}-(t-\Delta)\gamma_{2,B}\gamma_{3,A}+ ...]\nonumber\\
   &+ i[-(t+\Delta)\gamma_{1,B}\gamma_{2,A}+(t-\Delta)\gamma_{2,A}\gamma_{3,B}- ...].
   \label{decoupled}
 \end{align}
 Here we get one normalized zero mode, for $|t+\Delta|<|t-\Delta|$, from each of these chains that peaks at the end of the chain and decays exponentially away from it.
 
     Now let us define a $k$-space corresponding to this model of Hamiltonian \ref{ssh1d}.
For that we can define $c_{2i-1}=a_i$ and $c_{2i}=b_i$ and for periodic boundaries the Fourier transform looks like
  \begin{eqnarray}
    a_k&=\sqrt{\frac{2}{L}}\sum_{i=1}^{L/2} a_i e^{-ikx_{2i-1}}\nonumber\\
    b_k&=\sqrt{\frac{2}{L}}\sum_{i=1}^{L/2} b_i e^{-ikx_{2i}}
    \end{eqnarray}
 So the Hamiltonian becomes
  \begin{align}
    H&=\sum_{i=1}^{L/2}[(t+\Delta)a_i^\dagger b_i+(t-\Delta)b_i^\dagger a_{i+1} + h.c.]\nonumber\\
    &=\sum_{k}[(t+\Delta)a_k^\dagger b_ke^{ik}+(t-\Delta)b_k^\dagger a_ke^{ik} + h.c.]\nonumber\\
     &=\sum_{k}[(2t\cos k +2i\Delta \sin k)a_k^\dagger b_k + h.c.]\nonumber\\
    &=\sum_{k}\psi_k^\dagger[2t\cos k\sigma_x +2\Delta \sin k\sigma_y]\psi_k\nonumber\\
    &{=\sum_{k}\psi_k^\dagger H_k\psi_k}
    \label{ft}
  \end{align}
  with energy $\epsilon(k)=\pm2\sqrt{t^2\cos^2(k)+\Delta^2\sin^2(k)}$ which has a gap for $t,~\Delta\ne 0$. The gap closes for $k=0$ at $t=0$ (a fully staggered hopping model)   or for $k=\pm\pi/2$ at $\Delta=0$.
  {One can notice the similarity of this Hamiltonian with the 1D Kitaev chain Hamiltonian\cite{kitaev2} with modulation $\Delta$ acting as superconductor pairing potential there.} In Eq.\ref{ft},  $\psi_k=(a_k,b_k)^T$. we can give this an unitary rotation to $\psi_k\rightarrow(a_k,b_ke^{ik})^T$ and rewrite the Hamiltonian with $H_k=[(t+\Delta)+(t-\Delta)\cos 2k]\sigma_x +(t-\Delta) \sin 2k\sigma_y$. The eigenvector corresponding to negative energy becomes
  \begin{displaymath}
    \psi_k=\frac{1}{\sqrt{2}}\left(\begin{array}{c}
    e^{-i\phi_k}\\
    -1\end{array}\right)
  \end{displaymath}
  where $\phi_k= \tan^{-1}(d_y/d_x)$ with $d_y=(t+\Delta)+(t-\Delta)\cos(2k),~d_x=(t-\Delta)\sin(2k)$.
  Notice that as we move around the Brillouin zone, the state vector forms a circle along the equator of the Bloch sphere which never touches the origin, the gapless point for this model (for $t,~\Delta\ne0$). But the closed loop in $d_x-d_y$ plane can surround or not the origin, depending on the sign of $\Delta$ (assuming $t>0$). So $\Delta=0$ implies a gap closing and opening Lifshitz topological quantum phase transition. Keeping in mind that the reduced Brillouin zone being halved due to two sublattice structure of this problem, one can calculate the Berry phase to be
  \begin{eqnarray}
    \gamma_B=\int_{-\pi/2}^{\pi/2}{\langle\psi_k|i\partial_k|\psi_k\rangle} dk=\frac{\pi}{2}[1-sgn(t*\Delta)]
    \label{berry}
  \end{eqnarray}
  which leads to $\gamma_B=\pi$ for $\Delta<0$ and zero otherwise (assuming $t>0$). {This $\gamma_B$ represents the topological invariant in this system and given a finite chain with open boundaries, here one can witness the topological end modes (see Fig.\ref{edge}) as a signature of the bulk-boundary correspondence. These topological Majorana zero modes are robust under disorder as long as the chiral symmetry of the system is preserved. Even for a chirality breaking weak on-site disorder, the end modes can be considered chiral for practical purposes\cite{disorder}. It is interesting to note here that a chirality preserving disorder like a domain wall created using variation of $\Delta$ can cause the two zero modes to be localized at one end of the chain and at the domain wall position\cite{disorder}.}

\begin{figure}
      \includegraphics[width=\linewidth]{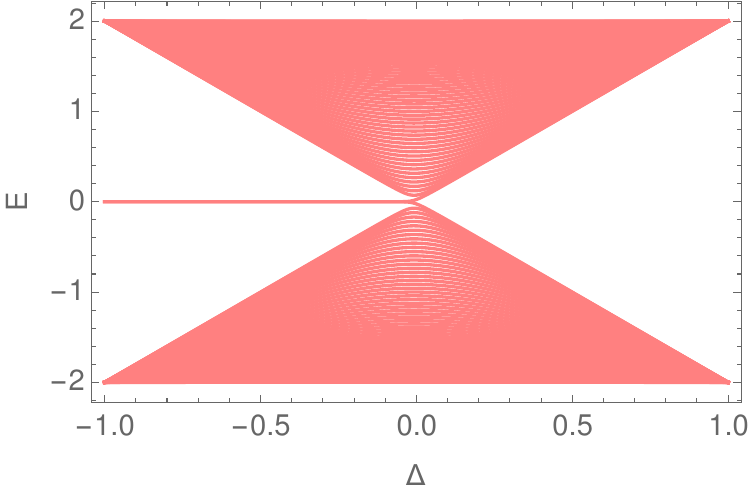}\\
      \vskip -1.65 in
      \hskip 2 in
      \includegraphics[width=.4\linewidth,height=.8 in]{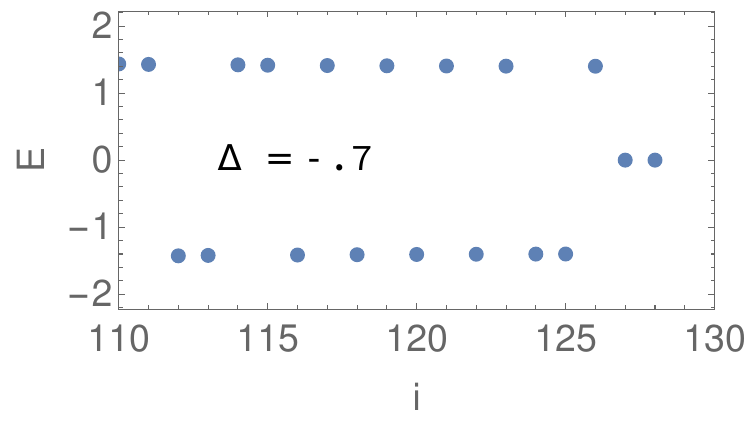}

\vskip .75 in
\caption{Numerical spectra of a SSH chain for L=128 as function of $\Delta/t$ for $\theta=\pi$. The inset shows energy of a few low energy states for $\Delta/t=-0.7$.} 
\label{cartoon}
    \end{figure}

\begin{figure}
      \vskip -.4 in
\begin{picture}(100,100)
  \put(-70,0){\includegraphics[width=.48\linewidth]{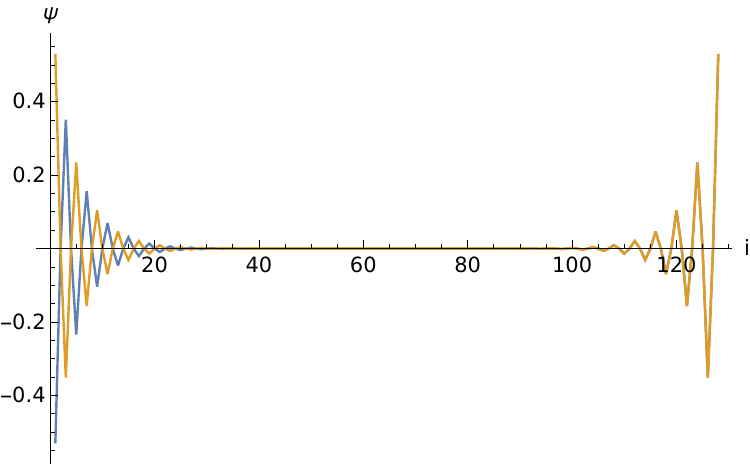}
    \includegraphics[width=.48\linewidth]{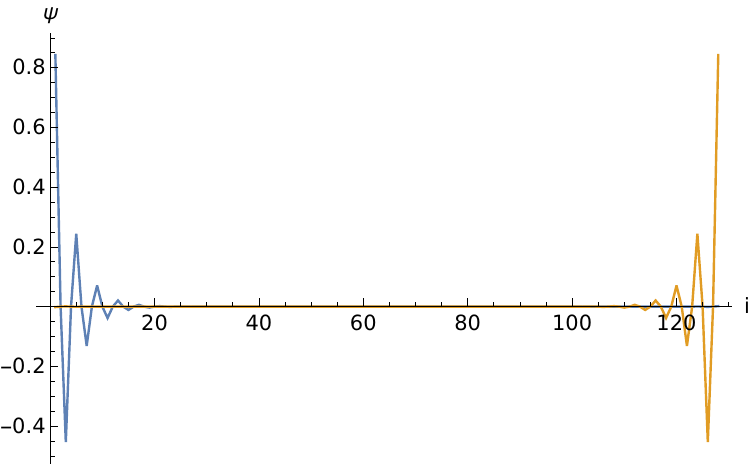}}
  \put(-10,60){(a)}
  \put(100,60){(b)}
  \end{picture}
\caption{End states of a SSH chain for L=128 with (a) $\Delta/t=-0.2$ and (b) $\Delta/t=-0.3$ for $\theta=\pi$.} 
\label{edge}
    \end{figure}

  Now we take a closer look at how the wavefunctions distribute over the chain. For the finite chain the spectra are shown Fig.\ref{cartoon}.
    The corresponding wavefunctions include, as mentioned earlier, two zero energy Majorana modes as end states.
    \begin{figure}[b]
      \vskip -.4 in
      \begin{picture}(100,100)
        \put(-70,0){\includegraphics[width=.48\linewidth]{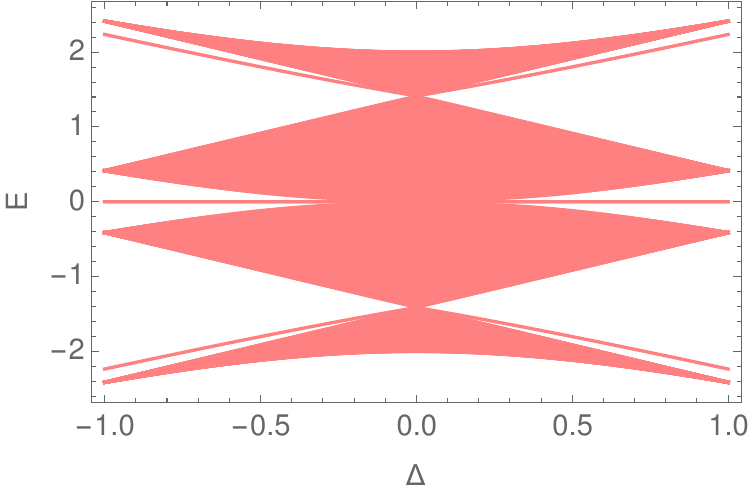}
          \includegraphics[width=.48\linewidth]{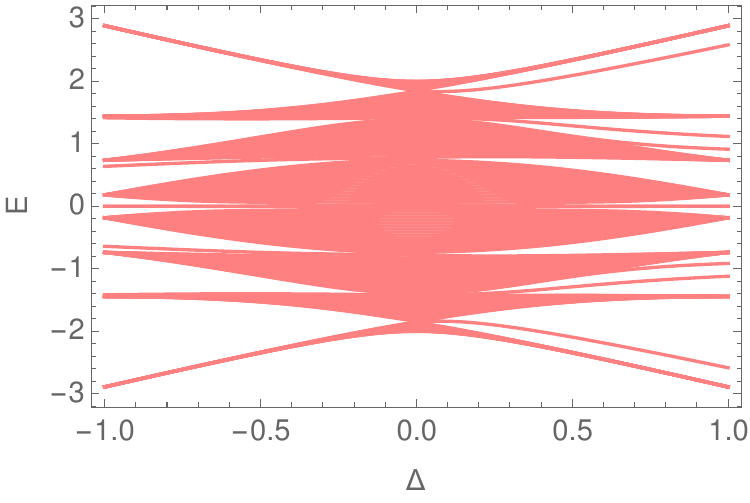}}
        \put(-10,45){(a)}
        \put(110,45){(b)}
      \end{picture}\\
      \vskip -.2 in
      \begin{picture}(100,100)
  \put(-70,0){\includegraphics[width=.48\linewidth]{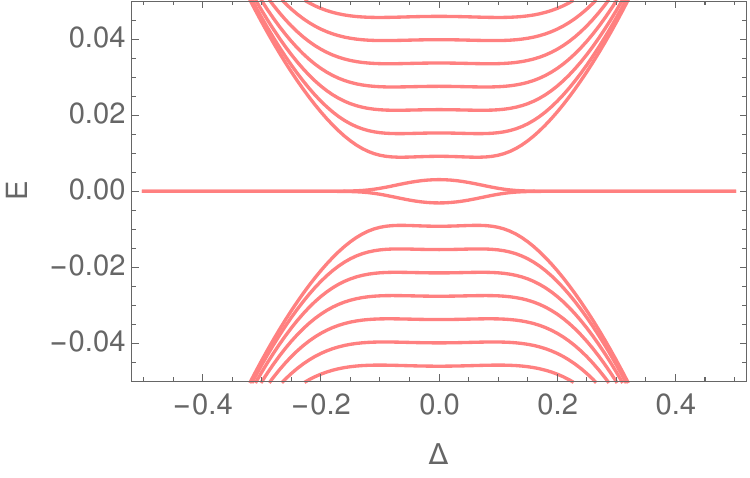}
  \includegraphics[width=.48\linewidth]{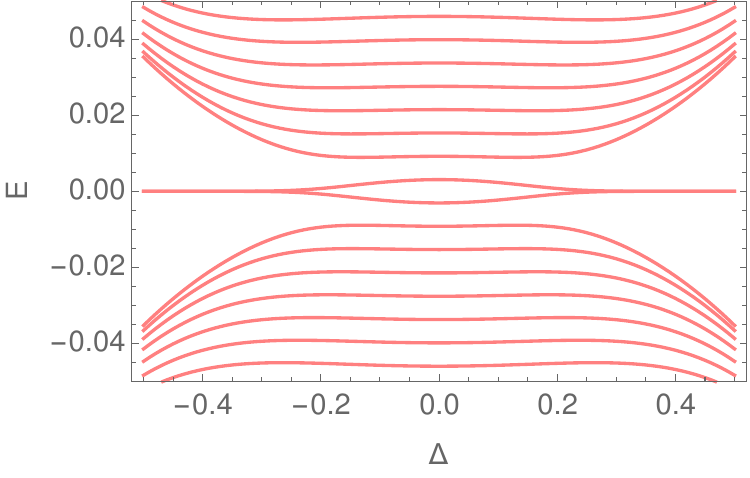}}
  \put(-5,55){(c)}
  \put(111,55){(d)}
        \end{picture}
\caption{{Numerical spectra of a SSH chain for $(a,b)$ L=128 and $(c,d)$ L=1024 as function of $\Delta/t$ for $(a,c)~\theta=\pi/2$ and $(b,d)~\pi/4$. The bottom panels show the same spectra as the top panels but zoomed in at low energies.}} 
\label{edge3}
\end{figure}

We see that for small -$\Delta$, symmetric and antisymmetric combinations are observed in the end state wavefunctions in small chains as a finite size effect. The end mode amplitudes decay away from the end as $|\frac{t+\Delta}{t-\Delta}|$ and hence for small $\Delta/t<0$ the modes get more delocalized leading to overlap and hybridization of the two end modes to give rise to symmetric and antisymmetric combinations of wavefunctions. However for large $\Delta/t<0$, the end-modes are more localized and appear in single end of the chain. Thus while varying $\Delta/t$ from negative towards zero value one comes across a {crossover} which occurs when low energy end-mode pairs hybridize to form symmetric and antisymmetric combinations along the length of the chain. Fig.\ref{edge} shows the pair of end-states for $\Delta/t=$-0.2 and -0.3 respectively which show symmetric and antisymmetric end-states for the former while single-end end-modes for the latter. In either case, the corresponding wave amplitudes show monotonous decay from the boundary in each sublattice.

Fig.\ref{edge3} shows the spectra for $\theta=\pi/2$ and $\pi/4$. Notice that for $\theta=\pi,~\pi/4$ the spectra changes with sign of $\Delta$ while for $\theta=\pi/2$ the spectra doesn't care for the sign. In fact one can find that for a chain of length $L=m*n$ and $\theta=\frac{2\pi}{2^n}$ ($m,n$ being integers), the spectra will (not) depend on the sign of $\Delta$ for all (even) odd $n$ values because the full set of hoppings in the chain differs with the sign of $\Delta$ only for an odd $n$. Fig.\ref{edge3} bottom panels highlight the low energy excitations including the MZM. {Notice that $\Delta=0$ implies the usual tight binding model without any topological phase in it.} And with $\theta=2\pi/2^n$ and $n\ge2$, unpaired MF exists for all nonzero $\Delta$ values (barring small $\Delta$'s where finite size effect causes symmetric and antisymmetric combination of end modes at small nonzero energies, like in Fig.\ref{edge3}(c)-(d)).

\begin{figure}
   \vskip -.4 in
   \begin{picture}(100,100)
     \put(-70,0){
  \includegraphics[width=.48\linewidth]{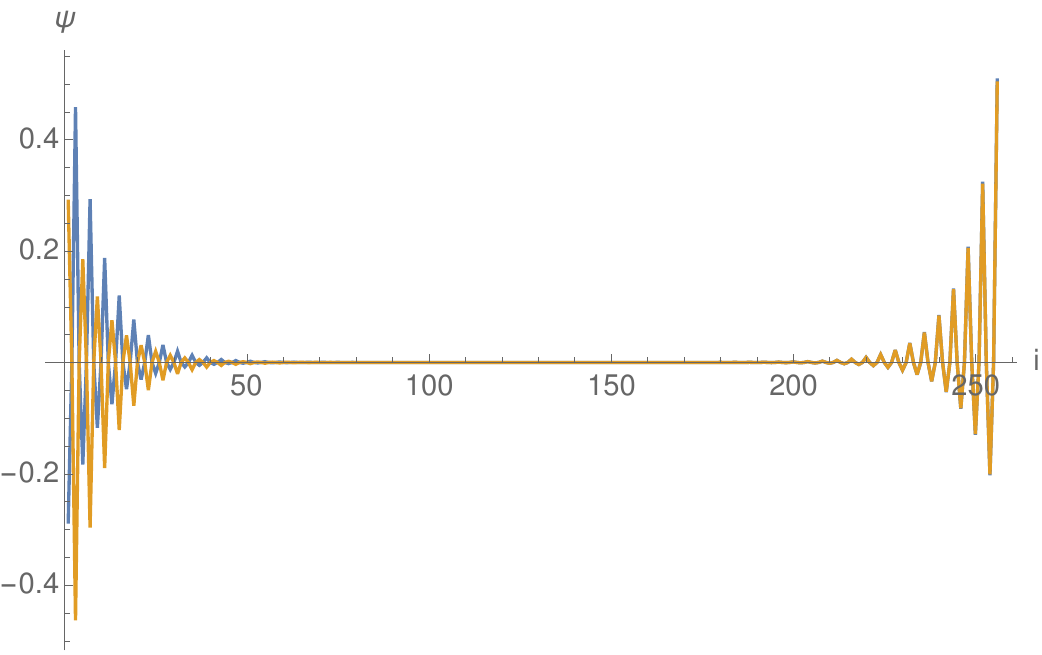}
  \includegraphics[width=.48\linewidth]{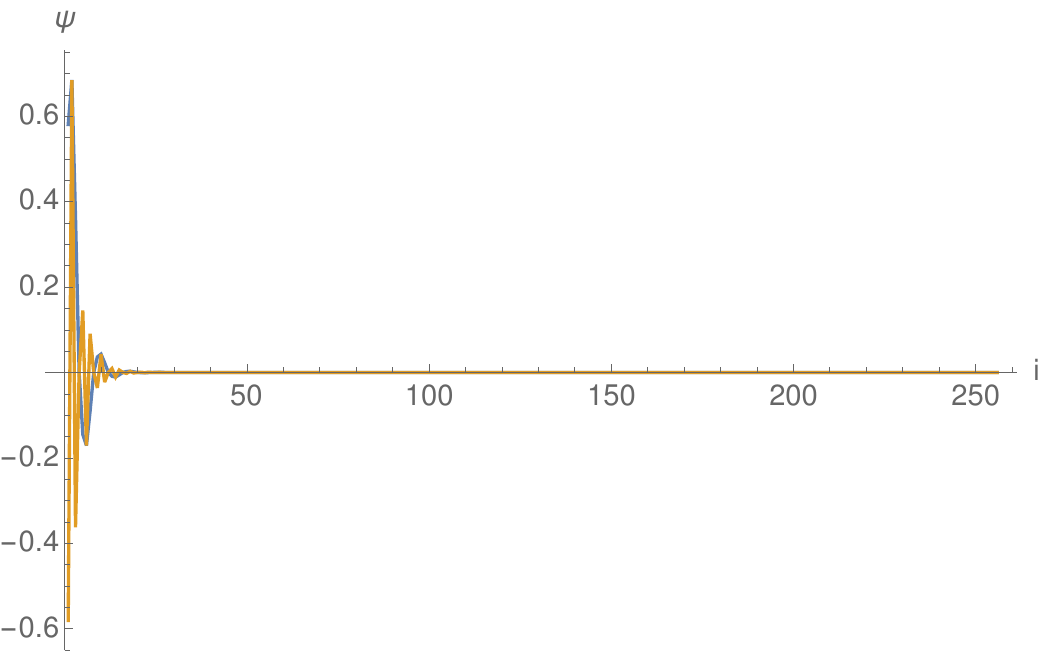}}
     \put(-10,60){(a)}
     \put(110,60){(b)}
   \end{picture}\\
   \vskip -.2 in
   \begin{picture}(100,100)
     \put(-70,0){
   \includegraphics[width=.48\linewidth]{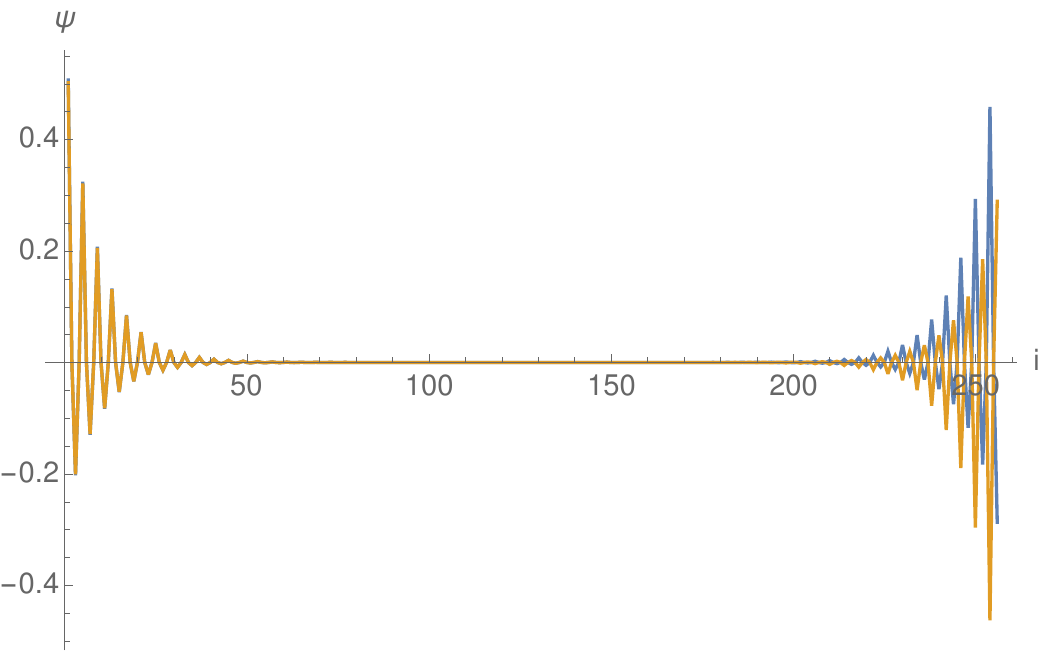}
   \includegraphics[width=.48\linewidth]{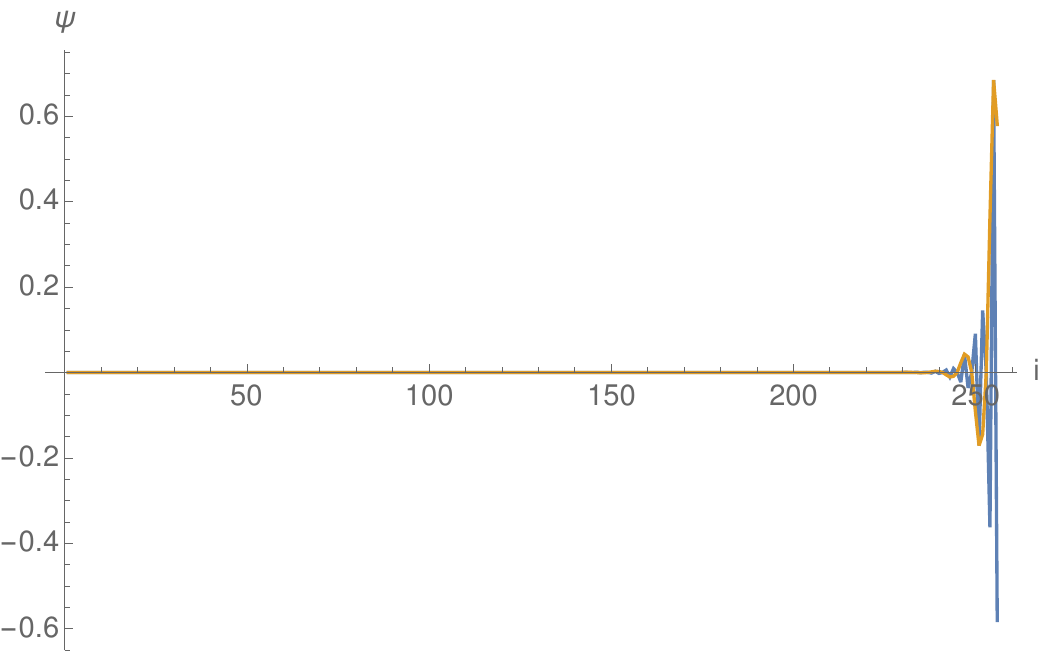}}
   \put(-10,60){(c)}
     \put(110,60){(d)}
   \end{picture}  
\caption{{Both-end (a,c) and single-end (b,d) end states of a SSH chain for L=256 with $\Delta/t=0.6$ (a,b) and $-0.6$ (c,d) for $\theta=\pi/2$.}} 
\label{edge2}
\end{figure}

 For $\theta=\pi/2$, we obtain a 4 sublattice structure and new gaps open up within the spectrum (see Fig.\ref{edge3}(a)). Out of the 4 in-gap states, two are found at nonzero energies (one positive and one negative) while the other two at zero energies. These zero energy states are Majorana bound states and appear as symmetric and antisymmetric combination of degenerate end modes or one-end modes depending on the size of $\Delta$ as well as $L$.  Fig.\ref{edge2} shows such end modes for $\Delta/t=\pm 0.6$ for $L=256$. The wave amplitudes of end states show monotonous decay from the boundary but with different rate of decay for different sublattices. Furthermore, we find the decay of the end modes away from the boundary to be slower as compared to the case with $\theta=\pi$. Similarly for $\theta=\pi/4$, we get a 8 sublattice structure yielding 6 nonzero in-gap end states (though not at completely identical energies for different signs of $\Delta/t$) and 2 zero energy Majorana states (see Fig.\ref{edge2}(b)). In essence, this is similar to getting more and more optical modes, in addition to the existing acoustic mode in the reduced zone picture of each Majorana spectrum as further branching of the Brillouin zone occurs due to further increase in periodicity of the hopping modulations.  Notice that in a chain with even number of sites, the end-states coming from the gap in the spectrum at nonzero energies are all single-end modes (see Fig.\ref{edge2} right panels) for they don't have any degenerate partner to form symmetric or antisymmetric combination and so these modes distinguish between boundaries.

 Generally for $\theta=2\pi/n$, we have $\delta_{i+1}=\Delta  cos(2\pi i/n)$ and the chain features a $n$ sublattice structure. The system is described by a $n\times n$ Hamiltonian matrix with $n$ number of eigenmodes. It's worth noting here that for $n=4$ or $\theta=\pi/2$, we get a $4\times4$ $H_k$ matrix with energy eigenvalue given as
 {$\epsilon(k)=\pm\sqrt{2t^2+\Delta^2\pm t\sqrt{2t^2+6\Delta^2+2(t^2-\Delta^2)\cos 4k}}$. Due to higher dimensionality, it is difficult to obtain the Berry phase (as given by the line integral in Eq.\ref{berry}) from the corresponding $4\times 1$ eigenvectors. It's rather easier, however, to obtain the Winding number (W) by enumerating the number of revolutions of $det[H_k]$ about the origin of complex plane as $k$ is traversed through the BZ\cite{lin,puel,gurarie,rmp}. For $\theta=\pi/2$, it turns out to be $W\ne0$ for $|t|>|\Delta|>0$ as we considered in this work\cite{comment}}. 
\\\\

\subsection{2D Model}

 Motivated by our findings in the SSH(like) chain, we also probe a 2D SSH model to see how the 2D edge states behave with tuning of $\Delta$ and $\theta$. One has the liberty to form the 2D SSH hoppings in different manner\cite{li,liu}.
We consider the symmetric hopping\cite{liu} in the Hamiltonian as
 \begin{align}
   H&=\sum_{i,j=1}^{L-1}(t+\delta_{i}) c_{i,j}^\dagger c_{i+1,j}+(t+\delta'_{j}) c_{i,j}^\dagger c_{i,j+1} + h.c.\nonumber\\
   &=\sum_{i,j=1}^{L-1}i(t+\delta_{i}) [\gamma_{i,j}^A\gamma_{i+1,j}^B+\gamma_{i+1,j}^A\gamma_{i,j}^B]\nonumber\\
   &+i(t+\delta'_{j}) [\gamma_{i,j}^A\gamma_{i,j+1}^B+\gamma_{i,j+1}^A\gamma_{i,j}^B]
   \label{2dssh}
 \end{align}
 where $\delta_i=\Delta cos[(i-1)\theta_x]$ and $\delta'_j=\Delta cos[(j-1)\theta_y]$ with $i,~j$ representing site index along $\hat x$ and $\hat y$ directions respectively and as previous, $\gamma_{i,j}^{A(B)}$ represents the Majorana operator at site $(i,j)$. It represents a finite $L\times L$ cluster with open boundary condition (OBC) at the edges.

 we first analyze the case for $\theta_x=\theta_y=\theta=\pi$.  In this case there will be in fact a 4-sublattice structure\cite{sk} and one can obtain the dispersions by Fourier transformation and diagonalization of the $4\times4$ matrix $H_k$ given by nonzero elements $H_k(1,2)=H_k(3,4)=t-\Delta+(t+\Delta)e^{-ik_x}$, $H_k(1,3)=H_k(2,4)=t-\Delta+(t+\Delta)e^{-ik_y}$ and their transpose conjugates\cite{liu}. One finds degenerate zero energy states along $(0,0)\rightarrow(\pi,\pi)$ nodal directions\cite{liu}. For a $L\times L$ finite square cluster, there are $L$ no. of zero-energy modes. Notice that for each site, there are now four nearest neighbors with direct hopping probabilities. Thus unlike in 1D, there are now four terms in the recursion relation for the zero modes ($i.e.,$ Eq.\ref{0mode}) and it does not, in general, lead to single-peaked localized states. The zero-modes are thus not necessarily corner or edge modes and hence not always single unpaired Majorana modes. Furthermore, all the nonzero energy modes are 4 fold degenerate due to 4-sublattice structure.
\begin{figure}
  \vskip -.1 in
  \begin{picture}(-150,70)
    \put(-210,0){
        \includegraphics[width=.335\linewidth]{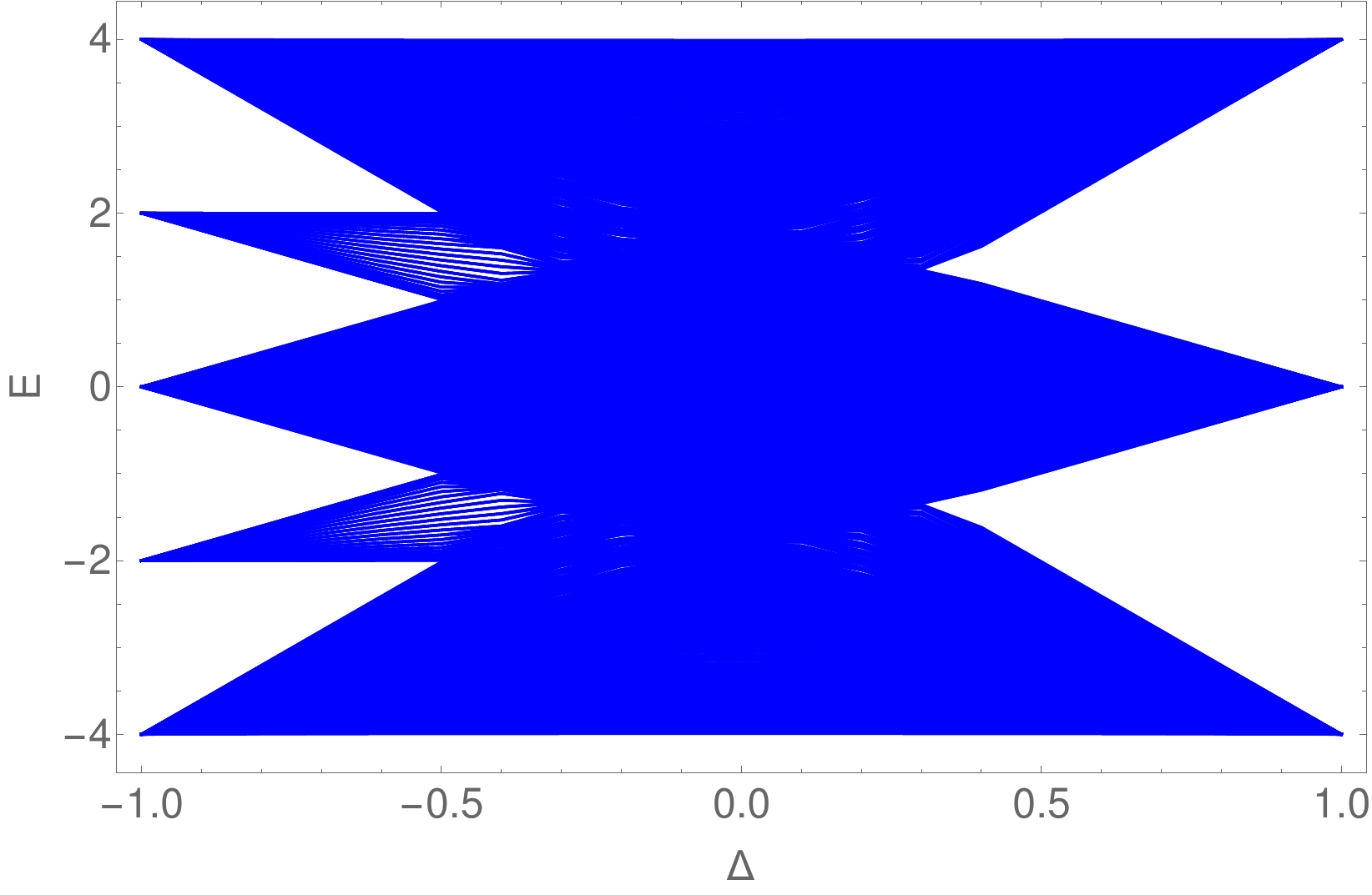}
      \includegraphics[width=.32\linewidth]{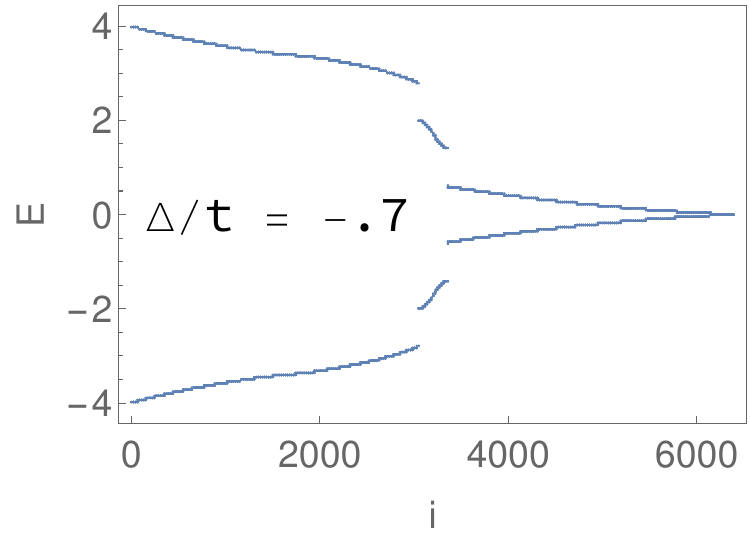}
      \includegraphics[width=.32\linewidth]{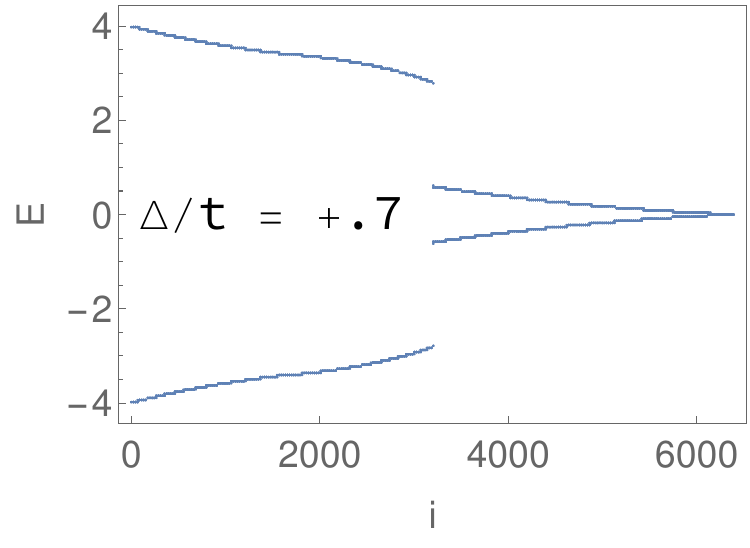}}
    \put(-170,70){(a)}
    \put(-90,70){(b)}
    \put(-10,70){(c)}
  \end{picture}\\
  \vskip -.25 in
  \begin{picture}(-150,100)
    \put(-210,0){
      \includegraphics[width=.32\linewidth]{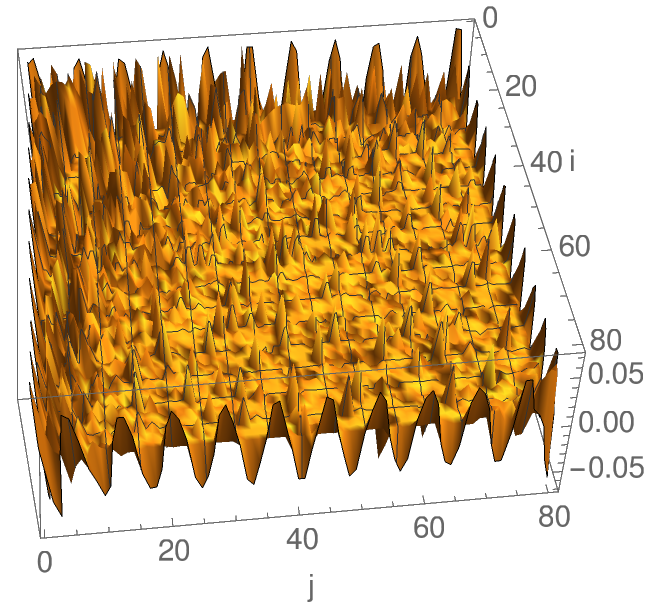}
      \includegraphics[width=.32\linewidth]{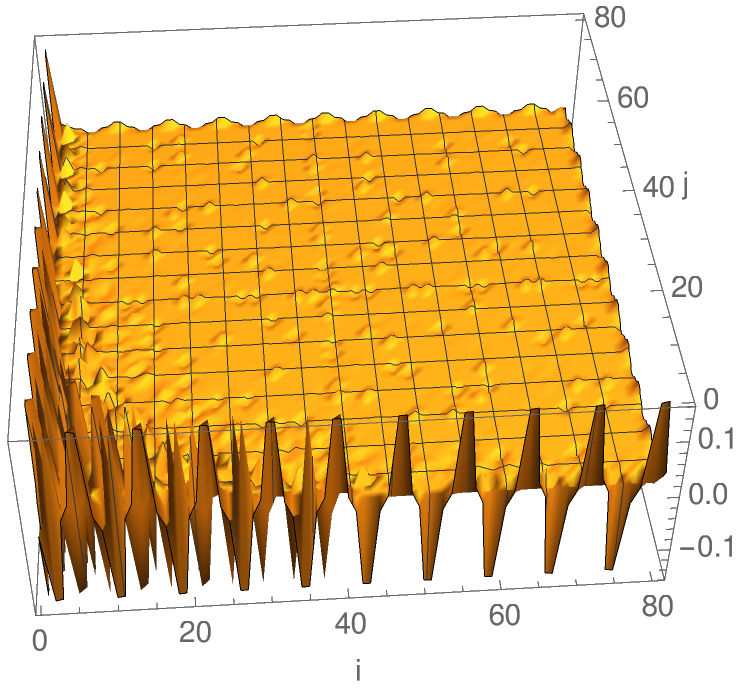}
      \includegraphics[width=.33\linewidth]{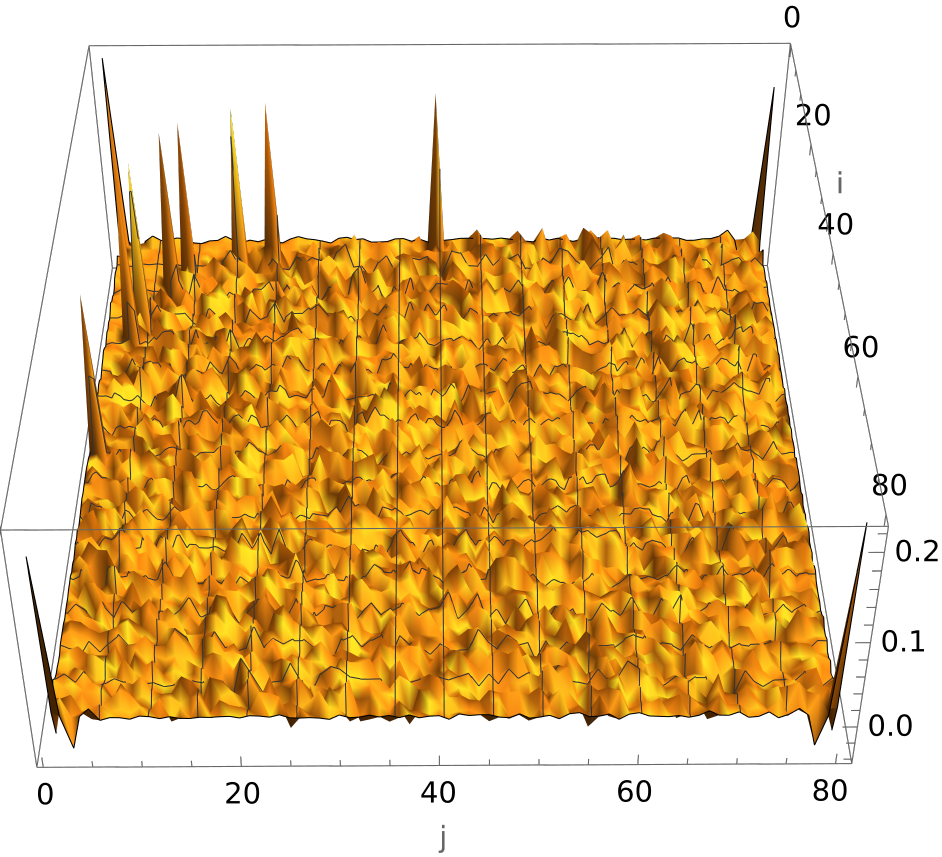}}
    \put(-170,70){(d)}
    \put(-90,70){(e)}
    \put(-10,70){(f)}
    \end{picture}
      
\caption{Dispersions of a 2D SSH model in a $80\times80$ square lattice (a) as a function of $\Delta$ for $\theta=\pi$. The same for $\Delta/t=$ $-0.7$ and $0.7$ are shown in panel (b) and (c) respectively. (d) and (e) show the typical {\bf nonzero energy in-gap edge states} for $\Delta/t=-0.2$ and $-0.7$ respectively while (f) shows {\bf zero energy corner state with satellite peaks} for $\Delta/t=-0.7$.} 
\label{2d-disp}
\end{figure}

 One should notice the existence of different available hopping pathways between sites in the 2D lattice. It also contributes to varieties of localized modes in a 2D SSH model. Among all the modes, one finds topological corner and/or edge states only for negative $\Delta/t$ values. Corner modes feature finite probability density at the corners of the lattice alone. Interestingly, localized modes with regularly located satellite peaks also appear as zero energy modes (Fig.\ref{2d-disp}(f)). Energy distribution among the eigenmodes for typical positive and negative values, $\Delta/t=\pm0.7$, are shown in Fig.\ref{2d-disp}(b)-(c). Notice that Fig.\ref{2d-disp}(a) clearly indicates gaps within the energy bandwidth for $\Delta/t\gtrsim0.3$ or $\Delta/t\lesssim-0.5$. But in addition to that, there are more small gaps with in-gap states within the spectrum. Edge modes are obtained as such in-gap states at nonzero energies for $\Delta/t<0$. Hence they are not Majorana modes. Even for small modulation $0>\Delta/t>-0.5$, where no gaps are readily discernible as in Fig.\ref{2d-disp}(a), edge states can be found at particular nonzero energies. The spectrum features zero energy degeneracies which makes the Berry curvature singular along the nodal direction which, for $\Delta/t<0$, sums up to nonzero Berry phases\cite{liu} producing topological behavior like edge excitations.
Interestingly for $\Delta/t<0$, there are few zero modes which, in addition to containing corner peaks, also feature a number of satellite peaks that are located following a pattern (at positions given by $xy=L$) within the lattice (Fig.\ref{2d-disp}(f)). That constitutes the beauty of Eq.\ref{0mode} in 2D. As mentioned earlier, the edge states at nonzero energies appear as in-gap states. Like in 1D, the decay of the edge modes away from edges are quicker for larger $\Delta$ and one can witness edge modes with wavefunction peaks at all the edges only for small $\Delta$ values (compare Fig.\ref{2d-disp} (d) and (e) and notice the similarity with the 1D case).

{Keeping edges only in one direction (say $\hat x$) while letting the other direction edge-less via periodic boundaries, one can obtain ribbon configurations which show some distinctive behavior. It does no more show any topological edge or corner modes at zero energies.} In fact, in the range of $-1<\Delta/t<1$ there is no zero energy state with PBC along $\hat y$. However, in-gap topological states at nonzero energies are obtained for $\Delta/t<0$.
Fig.\ref{2dssh} shows the typical edge modes, corner modes and extended modes for different $\Delta$ values for a square lattice with open boundary condition (OBC) along $\hat x$ direction and periodic boundary condition (PBC) along $\hat y$ direction.

\begin{figure}[t]
  \vskip -.4 in
  \begin{picture}(100,100)
    \put(-80,0){
   \includegraphics[width=.4\linewidth]{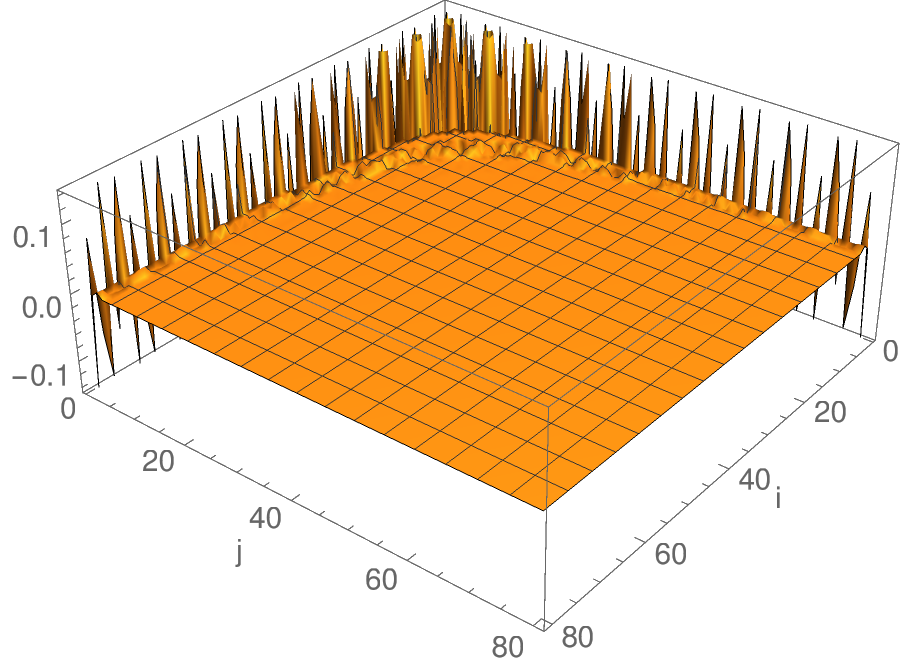}~~~~~~~~~~
   \includegraphics[width=.4\linewidth]{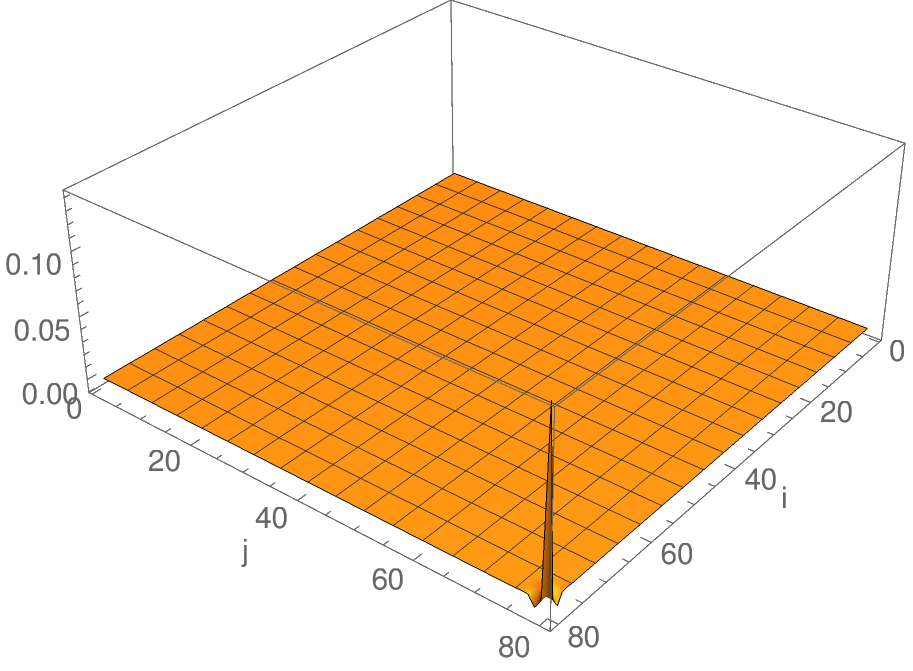}}
    \put(-70,5){(a)}
    \put(60,5){(b)}
    \end{picture}
  \vskip -.2 in
  \begin{picture}(100,100)
    \put(-80,0){
   \includegraphics[width=.4\linewidth]{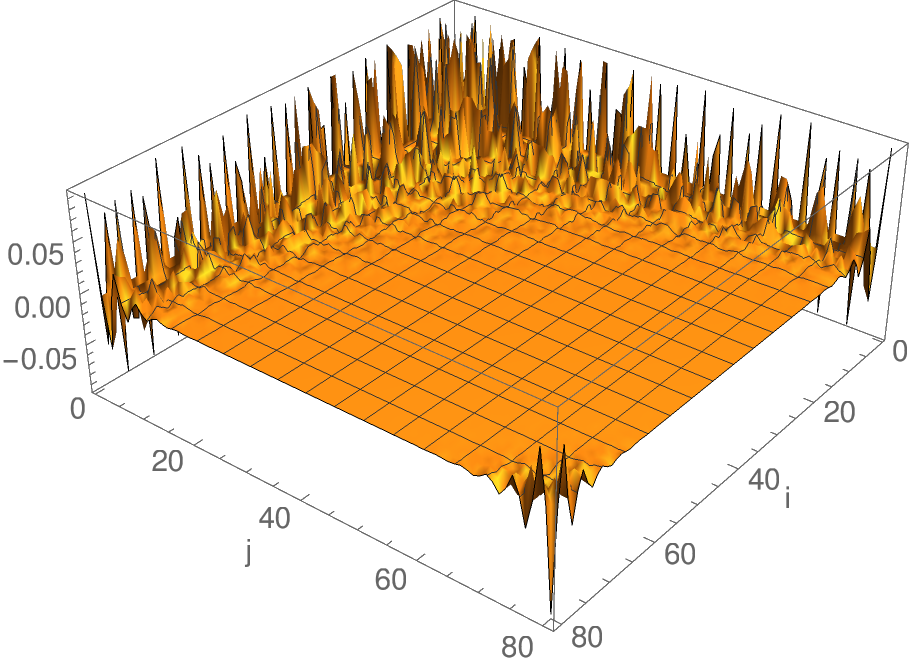}~~~~~~~~~~
   \includegraphics[width=.4\linewidth]{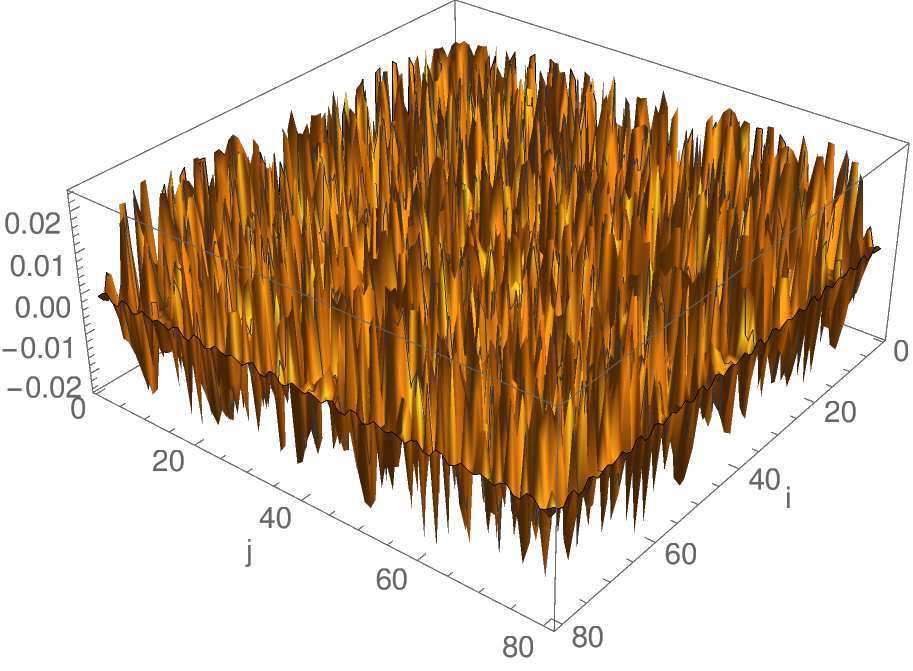}}
    \put(-70,5){(c)}
    \put(60,5){(d)}
    \end{picture}
\caption{Typical {\bf nonzero energy} edge, corner and delocalized extended eigenmodes of a 2D SSH model in a $80\times80$ square lattice with OBC (PBC) along $\hat x$ ($\hat y$) for $\theta=\pi$ and $\Delta/t$ = (a)-(b) -0.9, (c) -0.3 and (d) 0.3. } 
\label{2dssh}
\end{figure}

\begin{figure}[b]
   \vskip -.4 in
   \begin{picture}(100,100)
     \put(-70,0){
  \includegraphics[width=.48\linewidth]{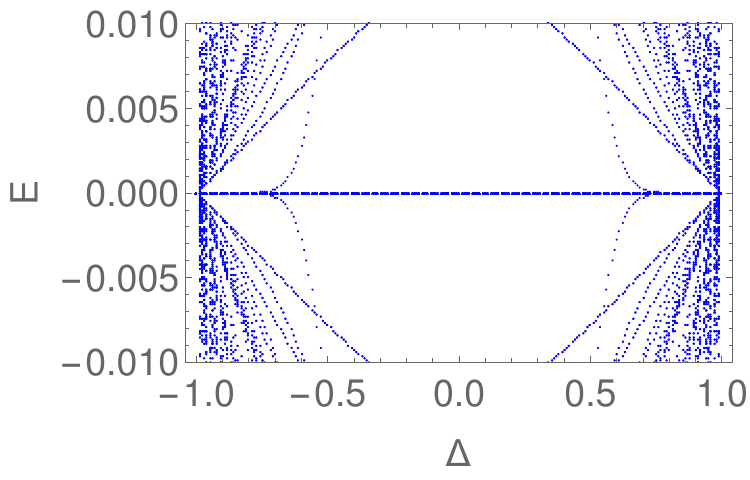}
  \includegraphics[width=.48\linewidth]{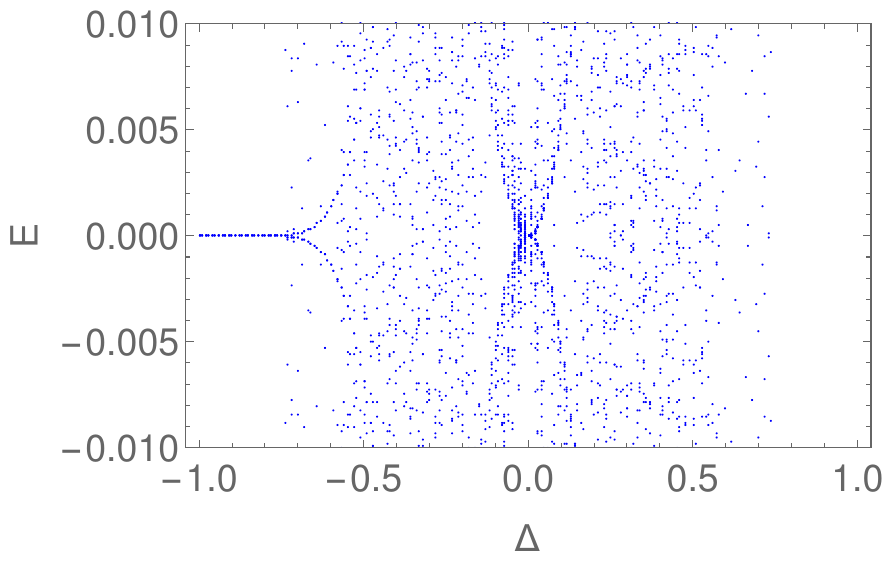}}
     \put(-0,60){(a)}
    \put(118,60){(b)}
  \end{picture} 
\caption{Low energy spectra of a 2D SSH model in a $80\times80$ lattice as function of $\Delta/t$ for (a) $\theta_x=\theta_y=\pi/2$ and (b) $\theta_x=\pi/2,~\theta_y=\pi$.} 
   \label{2d-per2per4}
\end{figure}

Next for symmetric variation of hopping with $\theta_x=\theta_y=\pi/2$, the spectrum becomes completely independent of the sign of $\Delta/t$ (like in 1D) and again there exists $L$ number of zero energy modes in a $L\times L$ cluster (with OBC along both directions). Fig.\ref{2d-per2per4}(a) gives the low energy spectrum of the same as a function of $\Delta/t$.

\begin{figure}[t]
  \vskip .4 in
   \begin{picture}(100,100)
     \put(-70,0){
   \includegraphics[width=.48\linewidth]{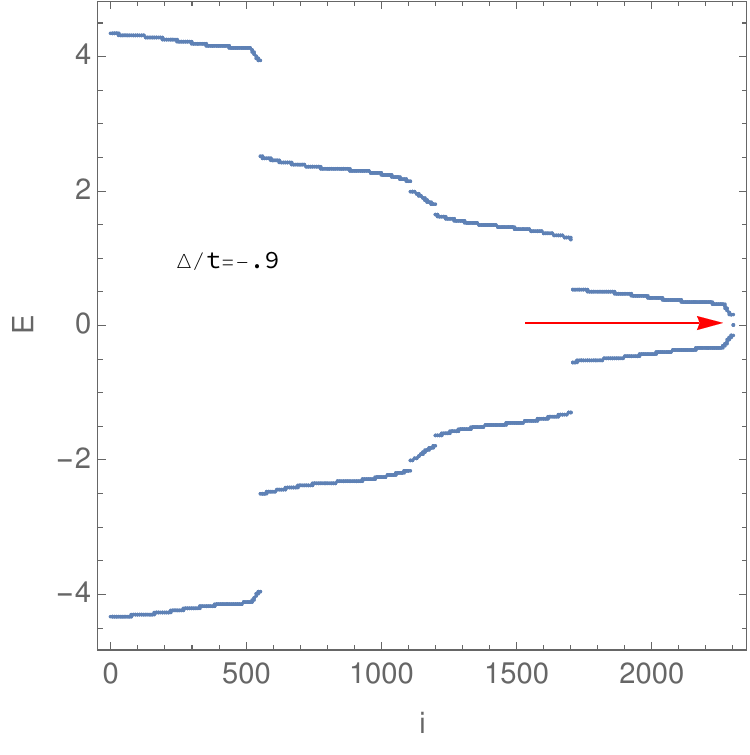}
   \includegraphics[width=.48\linewidth]{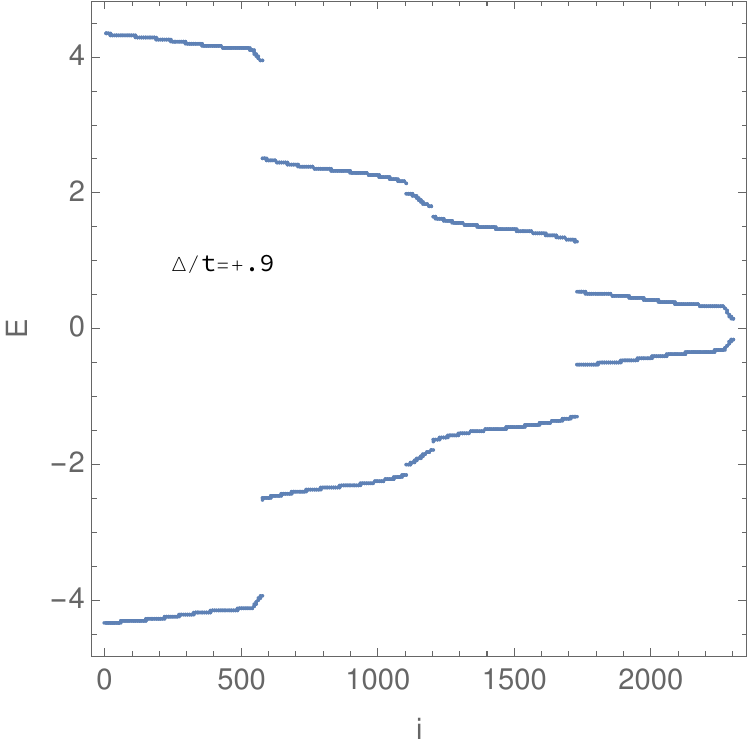}}
    \put(0,100){(a)}
    \put(130,100){(b)}
    \end{picture}
  \vskip -.2 in
  \begin{picture}(100,100)
    \put(-70,0){  
   \includegraphics[width=.48\linewidth]{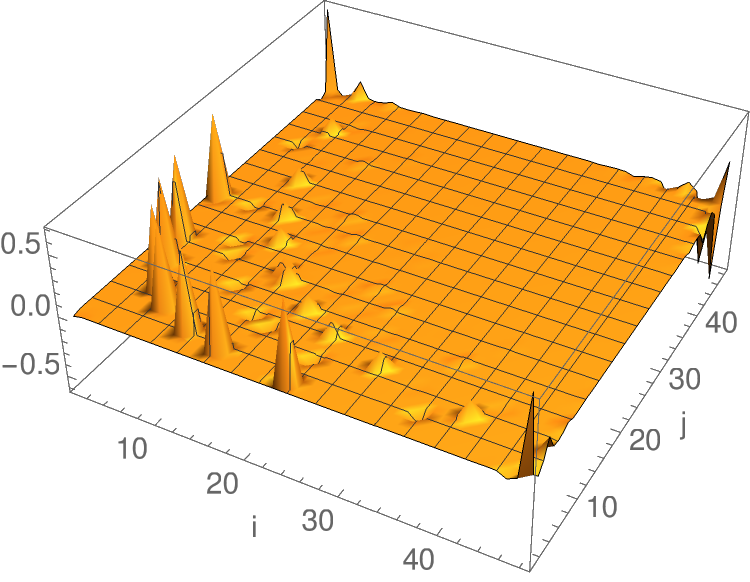}
   \includegraphics[width=.48\linewidth]{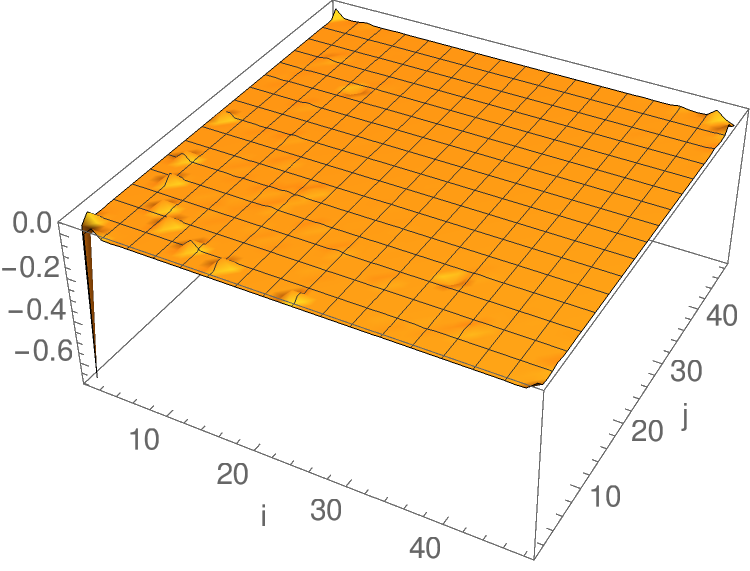}}
    \put(-0,50){(c)}
    \put(130,50){(d)}
  \end{picture}
\caption{Dispersion for $48\times48$ 2D SSH model with $\theta_x=\pi/2,~\theta_y=\pi$ for $\Delta/t$ = (a) -0.9 and (b) 0.9 respectively. Two typical {\bf zero energy modes} for $\Delta/t=-0.9$ are shown in panel (c) and (d).} 
\label{2d-per4}
 \end{figure}

 But if we consider different periodicity for the hopping modulation in the two directions, we get very different results. For a finite cluster with OBC in both directions, we find that the combination $\theta_x=\pi/2$ and $\theta_y=\pi$ shows interesting modification in the spectra and topology, namely the zero energy modes appears only for $\Delta/t\lesssim-0.8$ in a $80\times80$ cluster (Fig.\ref{2d-per2per4}(b)) and nonzero-energy in-gap states are obtained both for positive and negative $\Delta/t$ where edge states. Zero energy modes becomes rarer compared to the previous case with $\theta_x=\theta_y=\pi$ and in them one can find corner states. Fig.\ref{2d-per4} shows presence and absence of in gap corner states for $\Delta/t$=-0.9 and +0.9 respectively. Four zero energy states are obtained for $\Delta/t=-0.9$ (indicated via red arrow in Fig.\ref{2d-per4}(a)) which feature corner/edge excitations.

  \section{Discussion and Summary}
  In the present work, we have dealt with 1D and 2D SSH model with periodically modulated nearest neighbor hopping and discussed on the dispersions, spectral features and the end/edge states. Specifically, we focussed on topologically nontrivial regime which we found to vary with the hopping periodicity and its strength. Pair of MZM are obtained in all the different periodic hoppings that we studied in a 1D chain while end states as in-gap states at nonzero energies are obtained in addition, when the hopping periodicity becomes multiple of two lattice spacings. Due to availability of different hopping pathways, the features of topological states in a 2D SSH model are more rich where one obtains different types of edge states, corner states or states with discrete satellite peaks positioned non-randomly in a 2D lattice. Particular care was taken on studying the distribution of MZM in the systems. These modes indicate a sign dependence/independence of the modulation, based on the periodicity of the hopping. Interestingly, a different periodicity in the two directions indicate substantial decrease in the number of MZMs. The works on 2D SSH models are rather new  and we believe our work for the 2D system can be more established by precisely calculating topological indices like Berry or Zak phases\cite{liu} or recently proposed relative phase winding\cite{saptarshi} in each case of different periodicity and we plan to do that in a future communication.

{In practice one can engineer such hopping modulated models in cold atom systems in optical lattices\cite{xie}, in specially designed graphene nanoribbons\cite{gnr} or topological acoustic systems\cite{acoustic}} and examine the boundary modes to compare with results reported in this work. In order to understand these models more extensively, one can add further complexities such as studying varieties of spinful SSH models that incorporate Hubbard interactions between electrons\cite{stratos,tapan} and investigate the behavior of edge modes in presence of periodic modulation of hopping between the neighboring sites. {Similar study can also be imagined for an interacting Bosonic system with SSH-like dimerization\cite{bosonic}.} One can also play with both space periodic as well as time periodic hopping modulations in a SSH model and do a Floquet analysis\cite{torres} to understand the stroboscopic dynamics of such system and to get a complete phase diagram of the same. {Interestingly, topological plasmonic chains for nanoparticles, that act like a plasmonic analogue of SSH model, can show long range coupling between end modes due to retardation and radiative damping effect. It breaks chiral symmetry yet maintaining the topological edge states enabling radiative transport even via the localized edge modes\cite{plasmon1,plasmon2}. Keeping up with such exotic findings, it will be a good idea to examine these systems under additional periodic variation in the bonds/hoppings.}  Lastly, we may add here that a SSH Y-junction with braiding of defects and zero modes have been proposed to provide topologically protected quantum gates\cite{boross}. In this respect, it is worth trying to explore how a periodic modulation of hopping in a SSH chain can contribute to the quantum information processing.
 
\section*{Acknowledgements}
SK thanks K. Sengupta, D. Sinha, S. Mandal and P. Chatterjee for fruitful discussions. This work is financially supported by DST-SERB, Government of India via grant no. CRG/2022/002781.

\end{document}